# Mini−Review
# Reaction mechanism of hydrogen activation by frustrated Lewis pairs

Lei Liu,[a,†] Binit Lukose,[b] Pablo Jaque,[c] Bernd Ensing [d]

[a] Department of Physics & Earth Sciences, Jacobs University, Campus Ring 1, 28759 Bremen, Germany

[b] School of Electrical and Computer Engineering, Boston University, 02215 Boston, USA

[c] Departamento de Ciencias Químicas, Facultad de Ciencias Exactas, Universidad Andres Bello, Santiago, Chile

[d] Van't Hoff Institute for Molecular Sciences, University of Amsterdam, 1098 XH Amsterdam, The Netherlands

† Current address: Max Planck Institute for Polymer Research, Ackermannweg 10, 55128 Mainz, Germany

**Abstract**: Typically, a Lewis acid and a base react with each other and form classic acid−base adducts. The neutralization reaction is however prevented by the introduction of bulky substitutes and this interesting finding leads to a new concept called "frustrated Lewis pairs, FLPs". Since both reactivities of Lewis acids and bases are remained in the same systems, FLPs have been shown many important applications. One of them is hydrogen activation, which showed for the first time the use of a non−metal catalyst for that purpose. In this mini−review, we have summarized all important findings regarding the $H_2$ activation by FLPs. This includes pre−organisation of FLPs, reaction path for the activation, polarization of H−H bond and the factors affected the reactivity. In light of some recent developments, we aim to clarify the reaction mechanism for the $H_2$ actitation by FLPs, which has been under debate for decades since the first discovery of FLPs. We believe that this mini−review can be served as a guideline for the future fundamental studies and industrial applications.

**Introduction**

In 1923, Gilbert N. Lewis defined Lewis bases (LBs) as molecules which donate an electron pair, and Lewis acids (LAs) as molecules which receive an electron pair. When the lowest unoccupied molecular orbital (LUMO) of the LAs interacts with the lone electron−pair in the highest occupied molecular orbital (HOMO) of the LBs, a Lewis acid−base adduct is formed. This notion has become a primary axiom of chemistry, and a guiding principle in the understanding of chemical reactivity. However, there are several exceptions which deviate from this Lewis axiom. The first example was reported by Brown and co-workers[1] upon studying the reactions between pyridines and several boranes. They found that while most of the LAs and LBs reacted with each other and formed classical Lewis acid−base adducts, the mixing of α,α'–lutidine with trimethylboron B(CH$_3$)$_3$, resulted in no reaction even at −80 °C (Scheme **1a**).[2] Steric hindrance between the *o*-methyl groups in α,α'–lutidine and the methyl groups in B(CH$_3$)$_3$ was attributed to the behavior. This was the first study in which the term "steric hindrance" was used in the Lewis acid and base chemistry. Another exception was later reported by Wittig and Benz,[3] who did not observe formation of a classical Lewis adduct upon mixing of triphenylphosphine, PPh$_3$ (Lewis base, LB), and triphenylborane, BPh$_3$ (Lewis acid, LA). Instead, the addition of *o*−fluorobromobenzene to the above mixture yielded the *o*−phenylenebridged phosphonium−borate (Scheme **1b**). Similar phenomenon was observed by Tochtermann upon the addition of BPh$_3$ to butadiene monomer/triphenylmethane anion mixture. Instead of a classical adduct, a trapping product was obtained upon the addition of BPh$_3$ and Ph$_3$CNa to butadiene (Scheme **1c**).[4] Both researchers realized that the bulky Lewis pairs prevent the formation of classical Lewis acid−base adducts, and Tochtermann used the German term "antagonistisches Paar" to describe such a non−quenched Lewis pair.[4]

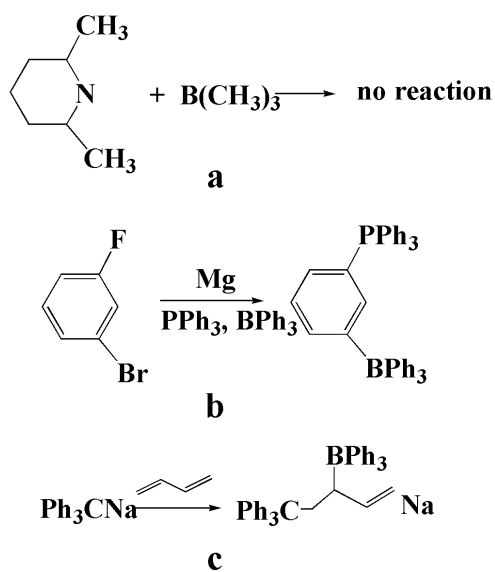

**Scheme 1**. Three examples of non−quenched Lewis pairs: (a) Addition of trimethylboron to α, α'−lutidine leads to no reaction. (b) Reaction of *o*−fluorobromobenzene with the mixture of triphenylphosphine and triphenylborane. (c) Addition of triphenylborane to the butadiene monomer/triphenylmethane anion mixture.

The term "frustrated Lewis pair" (FLP) was initially proposed by Stephan and co−workers in 2006.[5] In their study, they reported a covalently linked phosphino−borane $Mes_2P(C_6F_4)B(C_6F_5)_2$. In solution, this phosphino−borane exists as a monomer since both the B and P centers are sterically hindered, which precludes the dimerization or higher aggregation (Scheme **2**). On the other hand, this molecule contains both LA and LB fragments, and therefore it has been classified as an FLP. Importantly, it has been shown that the FLPs not only retain the typical reactivity of their individual components, but also show a cooperative acid−base character because of the two reactive sites. Thus, these bifunctional systems find application in various chemical processes, which were typically achieved by transition metal based catalysts. The applications include, activation of molecular hydrogen,[5,6] capture of $CO_2$,[7,8] $N_2O$,[9] $SO_2$,[10] and reduction of $CO_2$[11,12] imines.[13–15] This non−metallic and green catalytic system has attracted immense interest from researchers and FLP chemistry has become an important research area.[16] There are several review papers in the literature summarizing studies in the FLP chemistry.[17–26] In this mini−review, we focus on the studies which were aiming to address the nature of hydrogen activation by FLPs, that is, the process of heterolytic splitting of $H_2$ into a proton ($H^+$) and a hydride ($H^-$) in the presence of FLPs.

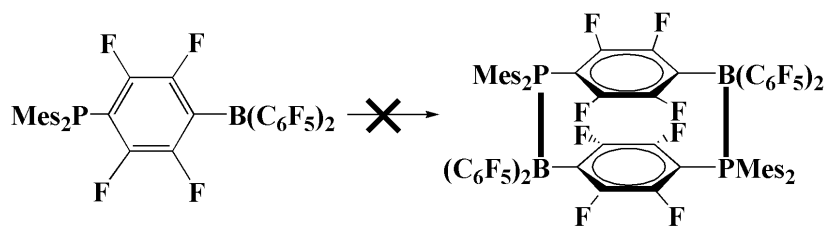

**Scheme 2**. Mes$_2$P(C$_6$F$_4$)B(C$_6$F$_5$)$_2$: the first example of a frustrated Lewis pair. Mes = 1,3,5-trimethylbenzene.

**Preorganization of FLPs.**

In general, there are two types of FLPs depending on their electronic structures. The first one is the intermolecular FLPs, in which the LA and LB centers are contained in two separated molecules (Scheme **3a**).[27] For this type of FLPs, it is assumed that when brought into contact in solution, the two individual components (LA and LB) associate into a loosely bound complex through secondary interactions mainly the London dispersions,[28,29] which then interacts with incoming small molecules, i.e. H$_2$, CO$_2$ and SO$_2$. However, early experimental attempts failed in the observation for the formation of such complexes. For example, the resonance signals in the NMR measurements of the FLP mixture were found to be identical to these of the individual components.[6] The yellow color of the $t$Bu$_3$P/B(C$_6$F$_5$)$_3$ mixture was thought to arise from the bound complex,[6,30] but a later experimental mechanistic study indicated that the color is characteristic of the $t$Bu$_2$PC$_6$F$_4$B(C$_6$F$_5$)$_2$ compound formed during the mixing of $t$Bu$_3$P and B(C$_6$F$_5$)$_3$.[31] Later, several advanced experimental techniques were employed to investigate the associations of the LA/LB. Typical examples are: Wiegand *et al.* found out that it is possible to distinguish between classic LA/LB adducts and FLPs through a solid−state NMR method.[32] By means of nuclear Overhauser effect spectroscopy (NOESY) measurements, Rocchigiani *et al.* found that the association of PMe$_3$/B(C$_6$F$_5$)$_3$ into a FLP is slightly endergonic ($\Delta G$ = 0.4 kcal mol$^{-1}$).[33] Nevertheless, associations of the LA/LB is remained unclear at a molecular level. The electronic structures of the loosely bound FLPs have been shown only theoretically, namely by density functional theory (DFT) calculations.[28,29,34–37] With electronic energy and solvent effects considered, the FLPs complexes are found to be more stable than the individual components, with an average association energy ($\Delta E_{solv}$) of −10 kcal mol$^{-1}$.[36] When entropy effects are included, the FLP complexes become less stable, and the computed Gibbs free energies ($\Delta G_{solv}$) of formation are around 5 kcal mol$^{-1}$,[36] which decrease to 2 kcal mol$^{-1}$ and to 0 kcal mol$^{-1}$ at accurate levels of theory.[28] Interestingly, a molecular dynamics (MD) simulation study proved that the association of two components indeed happened and lead to the formation of intermolecular FLPs, although the probability to find such FLP complex is extremely low, which is only 2 % of the total amount of phosphine and borane molecules used in the simulations.[38] Considering proper orientation of reactive centers, the concentration of reactive FLP complexes is even lower, which is only 0.5 %. Although the probability to form FLPs is low, $H_2$ can be activated by FLPs at a microscopic level. Based on the potential energy surface (PES) scan with respect to the distance between LA (boron atom, B) and LB (phosphorus atom, P) centers ($d_{PB}$), Liu *et al*. found that the optimal $d_{PB}$ for active intermolecular FLPs is between 3 Å and 5 Å.[36] When the substitutes on the reactive centers are too bulky and the repulsive interactions between two components are too large, we have a very long $d_{PB}$ (i.e. $d_{PB}$ is more than 5 Å for $t$Bu$_3$P/BMes$_3$).[36] This leads to inactive FLPs, which cannot be used for activation of $H_2$ or other small molecules.[6] On the other hand, when the substitutes are too small, then the LAs react with LBs, and form classic LA/LB adducts (i.e. Me$_3$P/B(C$_6$F$_5$)$_3$, Me$_3$P/B(C$_6$F$_4$H)$_3$ and TMP/BH(C$_6$F$_5$)$_2$).[6, 39, 40] It is important to point out that the shape of the PES is significant as well. In the cases of flat PESs, the two components can easily move without overcoming any energy barrier, and they may also form active FLPs with the minimum $d_{PB}$ not in the optimal range (3 Å < $d_{PB}$ < 5 Å).

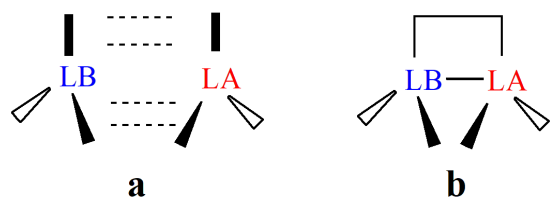

**Scheme 3**. Schematic representation of two types of FLPs: (a) intermolecular FLPs and (b) intramolecular FLPs. Notation used: LA for Lewis acid, LB for Lewis base.

The second type is the intramolecular FLPs, where the LA/LB components are in single molecules and are connected through C−C bridge (Scheme **3b**).[41] Unlike intermolecular FLPs, in which LA/LB centers are associated by secondary interactions with weak LA−LB interactions, the reactive centers in intramolecular FLPs are, however, intramolecularly connected to each other, with LA−LB distances similar to the classic LA/LB adducts. Taking the typical intramolecular FLP, $Mes_2P(CH_2)_2B(C_6F_5)_2$, as an example: the most stable structure is a four−membered heterocyclic phosphane–borane adduct, and the $d_{PB}$ is 2.2 Å.[42] To form an intramolecular FLPs and to activate $H_2$, we first need to open the closed ring structure. Certain amount of energy is needed to break the interactions between P and B, and increase the $d_{PB}$ from 2.2 Å to an ideal distance. DFT calculations found an open structure with $d_{PB}$ of 2.8 Å, and this structure lies 7 kcal mol$^{-1}$ above the closed structure on the PES. It is believed that such an open form is responsible for the $H_2$ activation.[42] Similar results have been found for a series of intramolecular FLPs where all open structures are 10 kcal mol$^{-1}$ less stable than the closed ones,[43] revealing the LA−LB disassociation of many intramolecular FLPs is thermodynamically mildly endothermic. Moreover, it was found out that the energy barriers for this opening−process (or pre−equilibration step) are also low.[44] To form intramolecular FLPs, the interactions between LA/LB centers should not be too strong. Otherwise high energy will be needed to form open structures, and this will reduce the reactivity, rendering FLPs inactive. In other words, the strength of LA/LB should be moderate which can be turned by the substitutes on both reactive centers. Besides, Erker *et al.* pointed out that the geometric parameters and conformational flexibility are also of great important.[45] Rigid LA/LB frameworks might reduce their reactivity towards $H_2$ and other small molecules.

**Reaction path for $H_2$ activation by FLPs.**

The first study on the reaction path for $H_2$ activation by FLPs has been reported by Stephan and co-workers.[5] The injection of $H_2$ into the solution of $Mes_2P(C_6F_4)B(C_6F_5)_2$ at 25 °C resulted in rapid generation of a zwitterionic salt, $[Mes_2PH]^+(C_6F_4)[BH(C_6F_5)_2]^-$, with color changing from orange to colorless. Heating this salt to 150 °C led to elimination of $H_2$ and regeneration of the phosphino-borane reactant (Scheme **2**). This remarkable finding represents the first transition metal free system that activates $H_2$ reversibly. In that study, the authors proposed two possible reaction paths for $H_2$ activation by $Mes_2P(C_6F_4)B(C_6F_5)_2$. One pathway is the addition of $H_2$ to the P−C bond, followed by the hydride migration from C to B. The other one is the addition of $H_2$ to the B−C bond, followed by the proton migration from C to P (Scheme **4**).

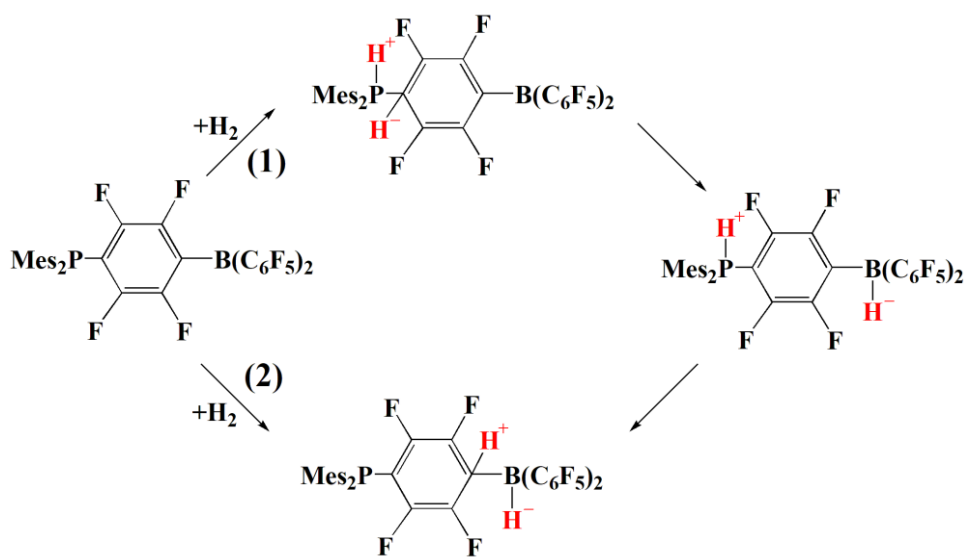

**Scheme 4**. Two possible reaction pathways for the reversible hydrogen activation by Mes$_2$P(C$_6$F$_4$)B(C$_6$F$_5$)$_2$: (1) hydride migration, and (2) proton migration. Mes = 1,3,5-trimethylbenzene.

Later, on the basis of DFT calculations, Guo *et al.*[46] proposed an alternative path for the reaction between H$_2$ and Mes$_2$P(C$_6$F$_4$)B(C$_6$F$_5$)$_2$. The authors explored and compared three possible pathways: (1) proton migration pathway; (2) hydride migration pathway; and (3) concerted Lewis acid−base pathway. Their findings indicate that the concerted Lewis acid−base pathway which involves π–π stacking of two Mes$_2$P(C$_6$F$_4$)B(C$_6$F$_5$)$_2$ molecules (Scheme **5a**) has a lower energy barrier in the rate−limiting step than that of the two migration mechanisms, which were proposed by the experimentalists (33.7 kcal mol$^{-1}$ versus 54.7 and 69.1 kcal mol$^{-1}$). In this concerted reaction path, two reactive centers (B and P) are from two Mes$_2$P(C$_6$F$_4$)B(C$_6$F$_5$)$_2$ molecules. As such, the H$_2$ activation proceeds through an intermolecular catalysis. The key step for such pathway is the dimerization of two Mes$_2$P(C$_6$F$_4$)B(C$_6$F$_5$)$_2$ molecules. The experimental findings of Welch *et al.*[47] regarding the crystal structure of the hydrogenated product, [(C$_4$H$_9$)$_2$PH]$^+$(C$_6$F$_4$)[BH(C$_6$F$_5$)$_2$]$^-$, is in support of the dimerization pathway. They found that the monomers pack in pairs, in a head−to−tail manner, with a short H$^{\delta+}\cdots$H$^{\delta-}$ distance of 2.6 Å (Scheme **5b**).

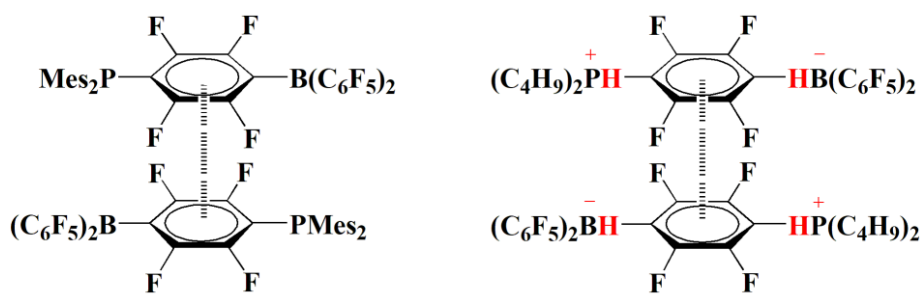

**Scheme 5**. The intramolecular π–π stacking of (a) two Mes$_2$P(C$_6$F$_4$)B(C$_6$F$_5$)$_2$ molecules, and (b) two [(C$_4$H$_9$)$_2$PH]$^+$(C$_6$F$_4$) [BH(C$_6$F$_5$)$_2$]$^-$ molecules. Mes = 1,3,5-trimethylbenzene.

Similar to that work, Liu *et al.*[48] reported computational studies on H$_2$ activation by a series of intramolecular FLPs, Mes$_2$PCHRCH$_2$B(C$_6$F$_5$)$_2$, with *R* = H, Me, Ph, and SiMe$_3$. These FLPs showed different reactivities towards H$_2$: in the cases of *R* = H and Me, the FLPs show H$_2$ activation, whereas in the cases of *R* = Ph and SiMe$_3$, the FLPs are inactive towards H$_2$.[49] The reaction paths that were known in the literature, which assumes a bimolecular reaction between the open FLP conformers and H$_2$,[34, 35] were not able to explain the experimentally observed different reactivities towards H$_2$. The authors therefore proposed an alternative reaction path, in which the key step is the stacking of two open FLP conformers into a dimer−like intermolecular FLP.[48] The dimerization of FLPs is possible for less bulky substituents, i.e. H and Me. However, this is not possible in the presence of bulky substituents like Ph or SiMe$_3$, and the reaction path becomes inaccessible because of the extra steric hindrance on the C−C bridge (**Scheme 6**).[48] Hence, this reaction path gives an adequate explanation for the selective reactivity towards H$_2$.

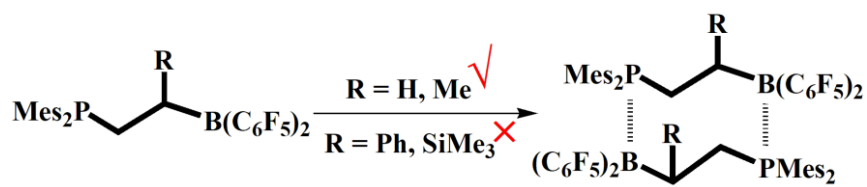

**Scheme 6**. The intramolecular π–π stacking of two Mes$_2$PCHRCH$_2$B(C$_6$F$_5$)$_2$ molecules, with *R* = H, Me, Ph and SiMe$_3$. Mes = 1,3,5-trimethylbenzene.

However, the reaction path that involves dimerization of FLPs is somehow in contrast to the experimental observations. The following examples illustrate that: Stephan *et al.*[5] performed kinetic studies using $^{31}$P ($^1$H) NMR spectroscopy to gain deeper insight into the reaction. The derived data showed that the reverse reaction

(losing of $H_2$) is first−order in terms of the concentration of $[Mes_2PH]^+(C_6F_4)[BH(C_6F_5)_2]^-$. Recent kinetic studies on other FLP systems also revealed that $H_2$ activation by a series of intramolecular FLPs is first−order in $[PH]^+/[BH]^-$ species.[50,51] Moreover, theoretical studies on the $H_2$ activation by FLPs indicates that the dimerization of FLPs is unnecessary to explain the reactivities of FLPs with $H_2$. On the ground of *ab* initio and DFT calculations, Rajeev *et al*. re−examined the $H_2$ activation by $Mes_2P(C_6F_4)B(C_6F_5)_2$ based on a simplified model system, $(CH_3)_2P(C_6F_4)B(CF_3)_2$.[52] The calculations showed that the reaction path, which involves a series of rearrangement reactions such as proton or hydride migration (**Scheme 4**) is both thermodynamically and kinetically reasonable. On the one hand, the computed reaction Gibbs free energies are in good agreement with the experimental observations, i.e. $H_2$ activation occurred at room temperatures and the reverse liberation process happened at an elevated temperature (i.e. 150 °C). On the other hand, the highest energy barrier along the lowest energy path was found to be around 30 kcal mol$^{-1}$, which is a reasonable value for the reactions happened at finite temperature.

Now, it is commonly accepted that reaction path for $H_2$ activation by FLPs involves one mole of FLPs, and without involving the dimerization of two FLP molecules. On the ground of static DFT calculations, one mole FLPs, one mole $H_2$ and one mole hydrogenation product ($[PH]^+[BH]^-$) are connected through only one transition state (TS), and $H_2$ activation by FLPs follows a concerted reaction path for both inter− and intramolecular FLPs (**Scheme 7a**). When $H_2$ interacts with FLPs, the H−H bond is polarized by the cooperation between LA/LB centers. After the heterolytic splitting of the H−H bond, the proton ($H^+$) and the hydride ($H^-$) are simultaneously captured by the LA and LB. Based on this reaction path model, the theoretically computed energy barrier for $H_2$ activation by the typical intramolecular FLP, $Mes_2P(CH_2)_3MeB(C_6F_5)_2$, is in perfect agreement with the experimentally measured values. DFT calculated energy barrier is 20.7 kcal mol$^{-1}$ while the corresponding experimental value is 22.3 kcal mol$^{-1}$. Moreover, the reaction shows a kinetic isotope effect of $k_{HH}/k_{DD}$ equal to 3.19 in the experiment while the computed value is 3.24.[51]

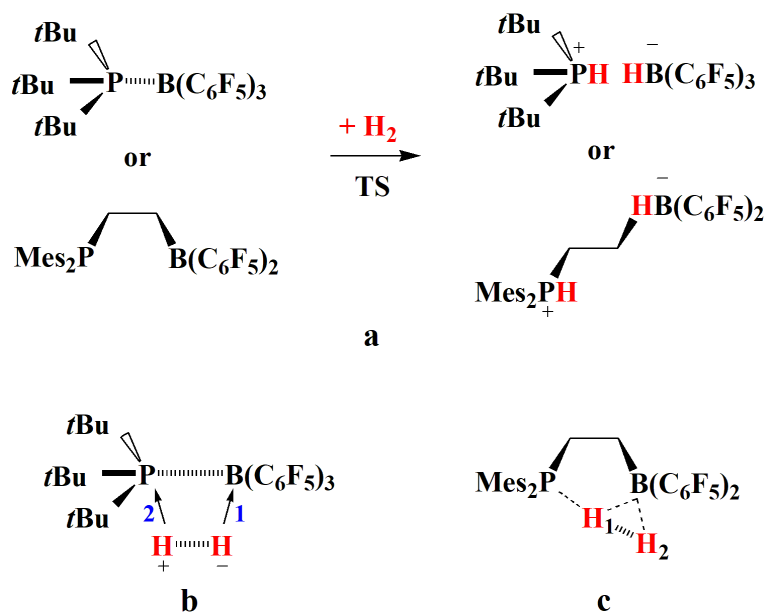

**Scheme 7.** (a) Concerted mechanism for H$_2$ activation by inter− and intramolecular FLPs obtained by DFT calculations. (b) Stepwise mechanism for H$_2$ activation by intermolecular FLP, obtained by AIMD. (c) TS for H$_2$ activation by intramolecular FLP obtained by DFT calculations. *t*Bu = *tert*−butyl. Mes = 1,3,5-trimethylbenzene.

Later, the detailed reaction path for H$_2$ activation by FLPs were investigated by *ab initio* molecular dynamics (AIMD) simulations.[53–56] The MD simulations confirmed the general profile which were uncovered by static DFT calculations: the dimerization of two FLP molecules is unnecessary, and the mole ratio of FLP to H$_2$ is equal to one. Interestingly, the MD simulations showed that the whole process of H$_2$ activation by FLPs contain two sub−steps (**Scheme 7b**). In other words, MD simulations revealed a step−wise mechanism: after polarization of H$_2$, proton (H$^+$) is first captured by Lewis acids (B), and hydride (H$^-$) is then captured by Lewis bases (P/N). In fact, such step−wise reaction path has also been indirectly suggested by static DFT calculations. In the structure of TS for H$_2$ activation by an intermolecular FLP, *t*Bu$_3$P/B(C$_6$F$_5$)$_3$, the only one imaginary frequency shows that the proton (H$^+$) has stronger interactions with Lewis acid center than that between the hydride (H$^-$) and base center. The B−H distance is considerable shorter than the P−H distance (1.68 Å versus 2.10 Å), and the computed covalent bond orders indicate that the B−H bond is formed earlier than the P−H bond (0.70 Å vs. 0.2 Å).[35] Similarly, DFT calculations showed that in the structure of TS for H$_2$ activation by an intramolecular FLP, Mes$_2$P(CH$_2$)$_3$MeB(C$_6$F$_5$)$_2$, H$_1$···H$_2$ overall has a side−on arrangement to B with an almost linear arrangement of the P⋯H$_1$⋯H$_2$ moiety (**Scheme 7c**). The only one imaginary frequency indicates that H$_2$ is activated at B center, and then P pulls the proton from activated hydrogen molecule.[51]

**Polarization of H−H bond.**

So far two DFT based models attempted to explain the polarization of H−H bonds have been proposed in the literature. The first one, electron transfer (ET) model, was proposed by Pápai et al.[34] who assumed that the LA/LB initially associates to form a weakly bound pair, which then interacts with H$_2$, and polarizes it in a concerted manner. A detailed molecular−orbital analysis[57] indicated that a simultaneous electron transfer from the lone pair of $t$Bu$_3$P to σ*(H$_2$) and from σ(H$_2$) to the empty orbital of B(C$_6$F$_5$)$_3$ occurs in a push−pull manner (Scheme **6a**), and results in weakening and subsequently heterolytic cleavage of the dihydrogen bond. In the transition state (TS) structure, H$_2$ occupies the reactive pocket of the P/B pair in an almost linear P–H–H–B arrangement with a slightly elongated H–H distance of 0.79 Å. The computed only one imaginary frequency corresponds to the stretching of the H−H bond and the formation of P−H and B−H bonds. The reaction barrier was associated with the energy cost for creating the orbital overlaps and distorting the individual donor and acceptor molecules.[57]

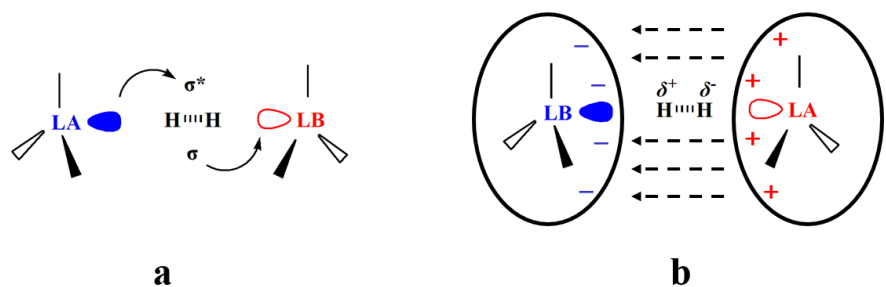

**Scheme 6**. Schematic representation of (a) the electron−transfer (ET) model of Papai et al.[34] (b) the electric−field (EF) model of Grimme et al.[35] Notation used: LA for Lewis acid, LB for Lewis base.

An alternative and simpler mechanistic picture of the FLP−driven H$_2$ activation was later proposed by Grimme and co-workers, which is the so−called electric field (EF) model.[35] It also assumes that a weakly bound pair is formed between the LA/LB through noncovalent secondary interactions (i.e. dispersion effects and hydrogen bonds). The incoming H$_2$ is then polarized and later split by the electric field created by the FLP (Scheme **6b**). The computed only one imaginary frequency corresponds to the stretching of the H−H bond and the entrance of the $H_2$ molecule into the FLP pocket. EF model suggested that the most uphill step is the entrance of the $H_2$ molecule into the FLP cavity, afterwards the reaction proceeds in a barrier−less manner, and there is no need to consider specific $H_2$/FLP orbital interactions.

It should be noted that there are some debates between these two models. On the one hand, EF model played in the $H_2$ activation was reinvestigated with several DFT and full CI methods.[58] All methods showed that activation barrier is strongly dependent on the electric field. Increasing the field strength drastically decreased the activation barrier, and the barrier even disappeared at strong experimentally accessible electric field. This investigation supported the original EF model, in which the key step is the polarization of hydrogen molecule. On the other hand, there are several studies insistent on the ET model. For example, investigating the reactions between simple Lewis pairs ($NH_3$ + $BX_3$, $X$ = H, F, Cl) and $H_2$, Camaioni *et al.*[59] found that the EF created by the $NH_3$/$BX_3$ pair indeed has a polarizing effect, but its contribution is so small that it cannot cleave the hydrogen molecule along. Later, the ET and EF models were deeply compared by examining a set of representative $H_2$ activation reactions.[60] The results showed that EF model has several drawbacks while the ET model could provide better understanding on the main features of the $H_2$ activation reactions. Recently, a MD simulation study, however, revealed that these two models are somehow complementary to each other.[56] For example, when $H_2$ is far away from the reactive centers with a distance longer than 2.5 Å, the H−H bond is polarized mainly through the electric field created by the FLPs. At such a large distance, the $H_2$ already gets polarized with an average H−H distance of 0.8 Å, and the electron density difference analysis does not show obvious electron transfer between $H_2$ and FLP molecules. When $H_2$ gets closer to the LA/LB centers with distance shorter than 2.5 Å, the electron density difference analysis clearly shows electron transfer from $H^−$ to B and P to $H^+$. In other words, the ET model is the most fitting model for the polarization of H−H bond in the short distance region. Nevertheless, the EF model may have contributions to the polarization of H−H bond in the region as well.

**Factors determining reactivity.**

Several reports have shown that intramolecular FLPs often show greater reactivity than their intermolecular counterparts.[61,62] This is most likely due to the fact that intramolecular FLPs do not have association issues since both LA/LB reactive centers are intramolecularly connected. The possibility to form intramolecular FLPs is much higher than that of the intermolecular FLPs in solution. For both types of FLPs, it has been realized by both experimentalists and theoreticians that their thermodynamic properties with $H_2$ is strongly affected by cumulative strength of the LA/LB, which can be described by the proton affinity and hydride affinity.[63] For example, it was found out that FLPs could show different reactivity with $H_2$ by changing the substituents on the LA/LB centers (Table **1**): $tBu_3P$ + $B(p\text{-}C_6F_4H)_3$ (FLP **1**) and $Mes_3P$ + $B(C_6F_5)_3$ (FLP **2**) showed non−reversible $H_2$ activation,[6, 64] $(o−C_6H_4Me)_3P$ + $B(p−C_6F_4H)_3$ (FLP **3**) showed reversible $H_2$ activation,[64] and $(C_6F_5)_3P$ + $B(C_6F_5)_3$ (FLP **4**) showed no reaction with $H_2$ in solution.[6] Based on DFT and *ab initio* calculations, Liu *et al.*[36] demonstrated that such different reactivities of FLP **1−4** can be explained by the cumulative strength of the LA/LB. The FLP **1** and FLP **2** contain either strong acid, $B(C_6F_5)_3$, or strong base, $tBu_3P$. Hence, the computed overall Gibbs free energies ($\Delta G$) are largely negative, indicating the zwitterionic products [([PH]$^+$[BH]$^−$)] are very stable and is hard to liberate $H_2$. The computed value of $\Delta G$ is close to zero for FLP **3** containing LA/LB with moderate strength. Therefore, this pair could, on the one hand, activate $H_2$. On the other hand, the final product could release $H_2$ upon heating up the solution. The strength of FLP **4** is so weak that it cannot split the H−H bond and reaction with $H_2$ showed a large positive $\Delta G$ values. Such relationship between reactivities and strength of FLPs has also been investigated by Tibor *et al.*[63] with quantum chemical calculations, by Rebecca *et al.*[65] and by Jiang *et al.*[66] with experimental techniques. Note that the cumulative acid–base strength not only determines the thermodynamic balance of the overall $H_2$ activation process, but also shows a systematic effect on the kinetics. According to the DFT calculations, Liu *et al.*[67] realized that increasing the strength of FLPs decreases the energy barriers for $H_2$ activation. In other words, the stronger FLPs have the lower energy barriers. Similar trend has also been found by Yepes *et al.* based on the theoretical investigation of $H_2$ activation by several intramolecular FLPs.[68]

**Table 1**. Experimentally examined combinations of Lewis acids (boranes) and Lewis bases

(phosphines) with reactivity towards H$_2$ activation. $t$Bu = $tert$−butyl. Mes = 1,3,5-trimethylbenzene.

|   | FLPs | Experimental results |
|---|------|----------------------|
| 1 | $t$Bu$_3$P + B($p$-C$_6$F$_4$H)$_3$ | nonreversible H$_2$ activation[6] |
| 2 | Mes$_3$P + B(C$_6$F$_5$)$_3$ | nonreversible H$_2$ activation[64] |
| 3 | ($o$−C$_6$H$_4$Me)$_3$P + B($p$−C$_6$F$_4$H)$_3$ | reversible H$_2$ activation[64] |
| 4 | (C$_6$F$_5$)$_3$P + B(C$_6$F$_5$)$_3$ | no reaction[6] |
| 5 | Mes$_2$P(CH$_2$)$_2$B(C$_6$F$_5$)$_2$ | H$_2$ activation[43] |
| 6 | Mes$_2$P(CH$_2$)$_3$B(C$_6$F$_5$)$_2$ | no reaction[43] |
| 7 | Mes$_2$P(CH$_2$)$_4$B(C$_6$F$_5$)$_2$ | H$_2$ activation[43] |

The thermodynamics of the H$_2$ activation by FLPs is also strongly affected by the environment, that is the organic solvent. Grimme *et al.*[20] pointed out that it is important to include the solvation free energies ($\delta G_{solv}$) when computing $\Delta G$, and it was shown that solvent contributions are around −10 kcal mol$^{-1}$ according to the COMOS−RS (conductor-like screening model for realistic solvents) calculations. A recent study was carried out to have a deeper insight into the reactivities of several intramolecular FLPs, Mes$_2$P(CH$_2$)$_n$B(C$_6$F$_5$)$_2$, towards H$_2$ (Table 1 FLPs **5**−**7**).[43] The DFT results showed that none of these FLP systems is kinetically problematic for H$_2$ activation since all computed energy barriers are under 15 kcal mol$^{-1}$. The authors revealed that the experimentally obtained different behaviors of FLPs **5**−**7** towards H$_2$ are mainly because of their thermodynamics. In detail, the overall H$_2$ activation by the inactive FLP is markedly endergonic (8.6 kcal mol$^{-1}$ for FLP **6**), while it is slightly endergonic (1.5 kcal mol$^{-1}$ for FLP **7**) or markedly exergonic (−7.5 kcal mol$^{-1}$ for FLP **5**) for the reactive FLPs. When $\Delta G$ was partitioned into gas−phase Gibbs free energies, $\Delta G_{gas}$, and solvation free energies, $\delta G_{solv}$, three important points have been made: 1) the $\delta G_{solv}$ are more than −13 kcal mol$^{-1}$ in all cases which is the biggest contributions (absolute values) when computing $\Delta G$. 2) all FLPs **5**−**7** cannot activate H$_2$ without the contributions from the solvent since all $\Delta G_{gas}$ are largely positive; 3) FLP **6** and FLP **7** have close values for $\Delta G_{gas}$ which are about 20 kcal mol$^{-1}$. FLP **7** is reactive with H$_2$ because of the large solvent contributions (−18.5 kcal mol$^{-1}$) while is only −13.6 kcal mol$^{-1}$ in the case of FLP **6**, rendering FLP **6** inactive. The importance of solvent is not only limited to H$_2$ activation, but also to other reactions of FLPs.[69–72] Interestingly, it has been found out that the solvation contributions can be replaced by crystal field when the reactivity of FLPs is transferred from solution to solid state (i.e. molecular crystal). Theoretical studies showed that the gas−phase reaction energies ($\Delta E$) are always more negative in solid state than that in both gas and solution phase for the same H$_2$ activation reactions.[73]

It is believed that the LA/LB work in a cooperated fashion, in both the formation of the FLPs and the splitting the H−H bond. However, the actual individual roles of the LA and LB are remained unclear: whether both components are important to the thermodynamics and kinetics, or one determines thermodynamics and the other one determines kinetics. To answer this fundamental question, Liu *et al.*[56] performed DFT based metadynamics simulations to investigate the reaction between H$_2$ and the typical FLP, *t*Bu$_3$P/B(C$_6$F$_5$)$_3$. Based on this advanced MD techniques, the authors obtained the continued free energy surface (FES) and the detailed reaction path for H$_2$ activation by *t*Bu$_3$P/B(C$_6$F$_5$)$_3$ pair (Figure **1**). The results showed that H$_2$ activation by *t*Bu$_3$P/B(C$_6$F$_5$)$_3$ pair mainly consists of two elementary steps: hydride (H$^−$) transfer to B and proton (H$^+$) transfer to P, which is quite consistent with the previous AIMD studies.[53–55] The important insight made by the authors is the explanation of the roles of LA and LB.[56] On the one hand, the results showed that the H$^−$ transfer to B has a much higher energy barrier than that of H$^+$ transfer to P (11.1 versus 0.8 kcal mol$^{−1}$). This finding indicates that H$^−$ transfer is the rate−determining step. On the other hand, H$^−$ transfer step is strongly endothermic ($\Delta G$ = 8.8 kcal mol$^{−1}$), and whole process becomes exothermic ($\Delta G$ = −3.3 kcal mol$^{−1}$) when H$^−$ transfers to P. In short, the kinetics is determined by LA while the thermodynamics is determined by the LB along H$_2$ activation.

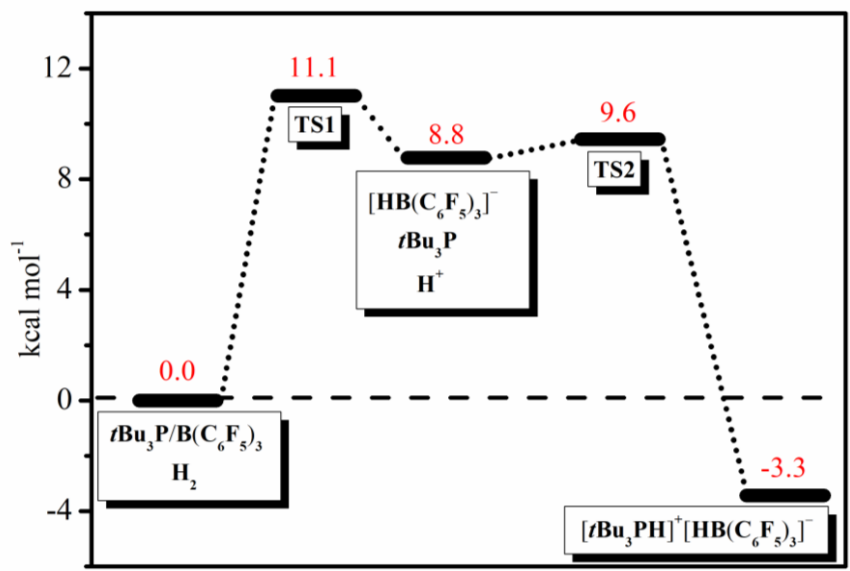

**Figure 1**. One−dimensional free energy surface (1D FES) for the $H_2$ activation by the FLP, $tBu_3P/B(C_6F_5)_3$, obtained by metadynamics simulations. $tBu$ = *tert*−butyl.

**Conclusions**

Due to its clean and renewable properties, hydrogen is considered as a sustainable energy resource and is largely used for hydrogenation reactions. Along its line of usage, activation of $H_2$ is the primary step, and FLPs have shown promising capacity for this purpose. The $H_2$ activation by FLPs generally follows several steps: the Lewis acid and base components first associates to form a FLP with an optimal distance and orientations through either secondary interactions or C−C linkage. After that, the $H_2$ enters the cave of the FLPs, and gets heterolytically polarized by the interactions with two reactive centers in a step−wise manner. At last, the Lewis acid and base centers capture the hydride and proton, respectively, forming the zwitterionic product. Quite often, the stronger strength of FLPs, the smaller energy barrier and more efficient catalytic properties are. The stability of the final product again depends on the strength of FLPs. The overall kinetics is mainly affected by the strength of the LAs while the thermodynamics is more affected by the strength of the Lewis LBs. In addition, the solvation effect is also important for the $H_2$ activation since the final hydrogenation product is often stabilized by the organic solvent. In this mini−review, we aim to address all important aspects concerning the $H_2$ activation by FLPs. We believe that this will help in the continuing research of FLPs, particularly that are focused on their industrial applications.

**Notes**

The authors declare no conflict of interest.